\documentclass[12pt]{article}
\usepackage{graphicx,amssymb}

\makeatletter

\def\paragraph{\@startsection{paragraph}{4}{\z@}{+2.00ex plus
 +1ex minus +.2ex}{1.5ex plus .2ex}{\it\normalsize}}

\def\section{\@startsection {section}{1}{\z@}{+3.0ex plus +1ex minus
  +.2ex}{2.3ex plus .2ex}{\normalsize\bf\boldmath}}
\def\subsection{\@startsection{subsection}{2}{\z@}{+2.5ex plus +1ex
minus +.2ex}{1.5ex plus .2ex}{\normalsize\bf\boldmath}}
\def\subsubsection{\@startsection{subsubsection}{3}{\z@}{+3.25ex plus
 +1ex minus +.2ex}{1.5ex plus .2ex}{\normalsize\bf\boldmath}}

\expandafter\ifx\csname mathrm\endcsname\relax\def\mathrm#1{{\rm #1}}\fi

\newcounter{saveeqn}

\@addtoreset{equation}{section}

\begin{document}

\begin{flushright}
\begin{minipage}{.2\linewidth}
UB-HET-03-03
April~2003
\end{minipage}
\end{flushright}

\vspace*{0.25cm}

\begin{center}

{\Large \bf Measuring the $W$ Boson Mass at Hadron
Colliders\footnote{Invited talk presented at the Mini-Workshop {\sl
``Electroweak Precision Data and the Higgs Mass''}, DESY Zeuthen,
Germany, February~28 -- March~1, 2003, to appear in the Proceedings} }

\vspace*{1cm}

{\sc U.~Baur}

\vspace*{.5cm}

{\normalsize \it
Physics Department, State University of New York at Buffalo \\
Buffalo, NY 14260, USA}
\par
\end{center}
\vskip 1cm
\begin{center}
\bf Abstract
\end{center} 
{\it
We discuss the prospects for measuring the $W$ mass in
Run~II of the Tevatron and at the LHC. The basic techniques used to
measure $M_W$ are described and  
the statistical, theoretical and detector-related uncertainties are
discussed in detail.
}
\par
\vskip 1cm

\section{Introduction}

Measuring the $W$ mass, $M_W$, is an important objective for the
Tevatron experiments in Run~II. One hopes to measure $M_W$ with a
precision of $\pm 30$~MeV per experiment per channel~\cite{run2_proc}. 
Together with the 
anticipated precision of 2~GeV for the top quark mass
from Tevatron data~\cite{tev2000}, a determination of the $W$ mass with 
an uncertainty of 30~MeV would make it possible to predict the Higgs boson
mass, $M_H$, from one-loop electroweak corrections to $M_W$ with an
accuracy of  
approximately $30\%$~\cite{tev2000,bd,erler}. With the best fit central
value close to the LEP~II direct search lower bound of
$M_H>114.4$~GeV~\cite{lep}, and the rather poor $\chi^2$ which results
from a fit of the
current precision electroweak data to the Standard Model
(SM)~\cite{chano,martin}, this would bring considerable additional
pressure on the SM. 

This document is structured as follows. 
The basic techniques used to measure the $W$ mass and the expected
statistical and overall uncertainties are briefly 
discussed in Sec.~\ref{sec:techniques}. The detector-related and
theoretical uncertainties affecting the $W$ mass 
and width measurements are described in more detail 
in Sec.~\ref{sec:det_uncertainties} and 
Sec.~\ref{sec:theo_uncertainties}, respectively. Alternative methods
of measuring the $W$ mass at the Tevatron are  
discussed in section~\ref{sec:alternatives}. Section~\ref{sec:conclusions} 
concludes this document.

\section{$M_W$ Measurement from the $M_T$ Lineshape}
\label{sec:techniques}

\begin{figure}[t!]
   \begin{minipage}[b]{.475\linewidth}
    \hspace{-0.2truecm}
    \includegraphics[width=3.1in]{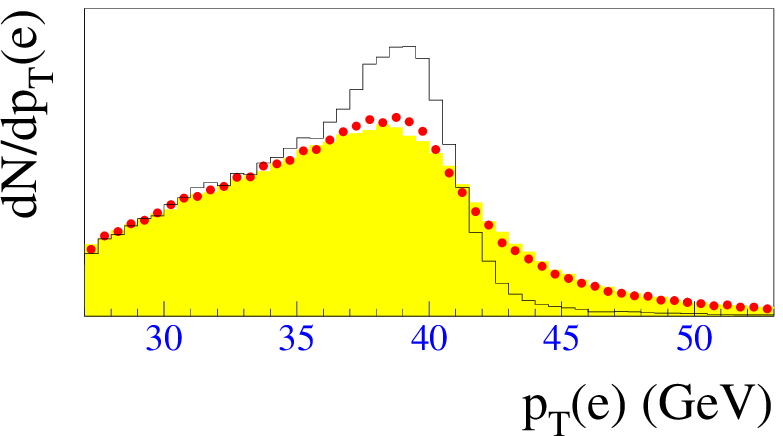}
\caption{The  effects of resolution and the finite $p_T^W$ on $p_T(e)$ in $W$
boson decay. The histogram shows $p_T^W$ without detector smearing and for
$p_T^W=0$. The dots include the effects of adding finite $p_T^W$, while the
shaded histogram includes the effects of detector resolutions. The effects are
calculated for the D\O\ Run~I detector resolutions.} 
\label{ptonly}
  \end{minipage}
   \hspace{0.7truecm}
   \begin{minipage}[b]{.475\linewidth}
    \hspace{-0.3truecm}
    \includegraphics[width=3.1in]{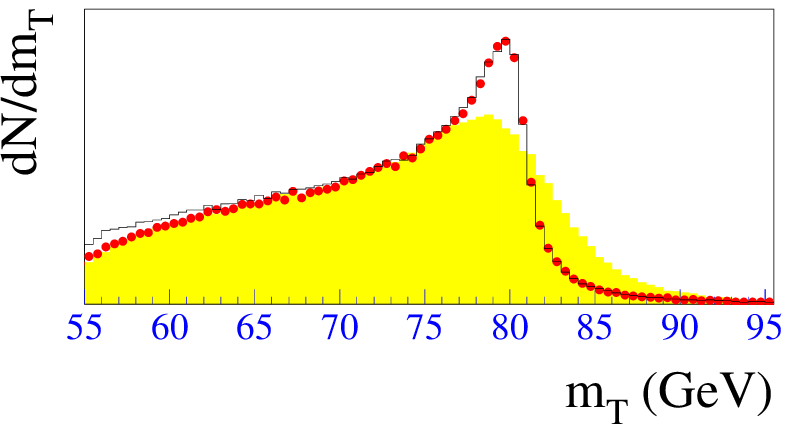}
\caption{The effects of resolution and the finite $p_T^W$ on $M_T$ 
in $W\to e\nu$. The histogram shows $M_T$ without detector 
smearing and for $p_T^W=0$. The dots include the effects of adding 
finite $p_T^W$, while the shaded histogram includes the effects of 
detector resolutions. The effects are calculated for the D\O\ Run~I 
detector resolutions.}
\label{mtonly}
    \vspace{0.1truecm}
  \end{minipage}
\end{figure}

The determination of $M_W$ depends on the two body nature of the $W$
decay, $W\to \ell \nu_\ell$. The kinematical Jacobian peak and
sharp edge at the value of $M_W/2$ is easily observed in the measurement of
the transverse momentum ($p_T$) of either of the leptons. However, the 
Jacobian peak in the $p_T$ distribution is smeared out by finite $W$
width effects, detector resolution effects, and, in particular, the
effect of the finite transverse momentum of the $W$ boson,
$p_T^W$. Figure~\ref{ptonly} 
shows a calculation of the $p_T(e)$ distribution (unsmeared) with
$p_T^W=0$; the effect of 
finite $p_T^W$; and the inclusion of detector smearing effects. It is
apparent that $p_T(e)$ is very sensitive to the transverse motion of the
$W$ boson.

Historically, precise understanding of $p_T^W$ has been lacking,
although it is currently modeled by measurable parameters through the
resummation formalism of Collins, Soper, and Sterman~\cite{collins}. For
this reason, the transverse mass, $M_T$, was suggested~\cite{smith} and
has been the traditional measurable. It is defined by 
\begin{equation}
M_T = \sqrt{2p_T(\ell) p_T(\nu) (1-\cos(\phi_{\ell,\nu}))},
\end{equation}
where $\phi_{\ell,\nu}$ is the angle between the charged lepton and the
neutrino in the transverse plane. The observables are the lepton transverse
energy or momentum $\vec{p}_T(\ell)$ and the non-lepton transverse
energy $\vec{u}$ (recoil transverse energy against the $W$), 
from which the neutrino momentum $\vec{p}_T(\nu)$
and the transverse mass $M_T$ are derived.  $M_T$ is invariant under transverse
boosts to first order in the velocity of the $W$ boson. 
Figure~\ref{mtonly} shows that the sensitivity of $M_T$  to $p_T^W$ is
indeed almost negligible. 
While considerably more stable to the phenomenology
of the production model, the requirement that the neutrino direction be
accurately measured leads to a set of experimental requirements which
are difficult in practice to control. 

\begin{figure}[t!]
\centering\leavevmode
\includegraphics[width=3.25in]{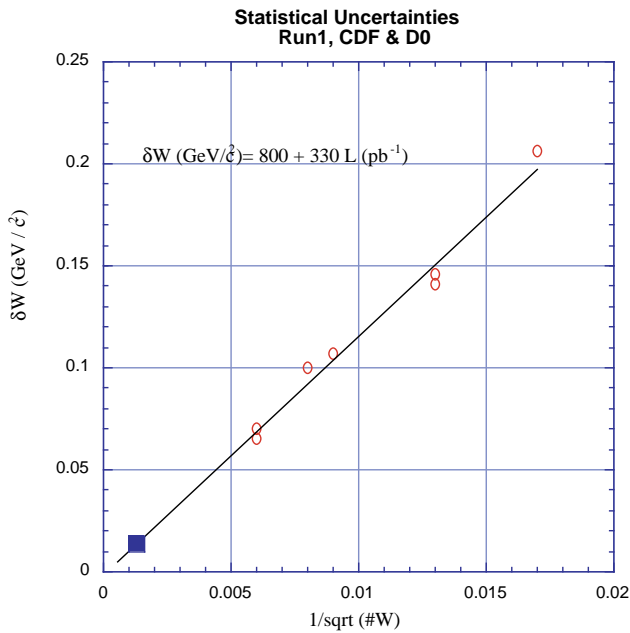}
\caption{Statistical uncertainties in Run~I $M_W$ measurements. Each circle
represents either a CDF or D\O\ measurement. The result of a straight line fit
is shown. The shaded box is the approximate extrapolation to a 2 fb$^{-1}$ 
Run~II result.}
\label{mw_stat}
\end{figure}

Both CDF and D\O\ have determined the $W$ boson mass using the
transverse mass approach. The combined $p\bar p$ collider result
is~\cite{martin} 
\begin{equation}
M_W=80.454\pm 0.059~{\rm GeV}.
\end{equation}
Presuming that Run~II is to deliver an integrated 
luminosity of 2~fb$^{-1}$, the statistical precision on $M_W$ can be 
estimated from the existing data. Figure~\ref{mw_stat} shows the  
statistical uncertainties in these measurements as a function of $1 
/\sqrt{N_W}$ ($N_W$ is the number of $W$ boson events), demonstrating a
predictable extrapolation to  
$N_W \simeq 700,000$ which corresponds to the expected Run~II dataset
per experiment per channel. The statistical 
uncertainty from this extrapolation is approximately 13~MeV. Taking into
account the detector related and theoretical uncertainties discussed in
more detail below, a total uncertainty of $\pm 30$~MeV for $M_W$ is
projected. The current CDF and D\O\ Run~II data samples are approximately
70~pb$^{-1}$~\cite{kotwal}. 

For integrated luminosities larger than 2~fb$^{-1}$, the systematic
errors rapidly become dominant. For 15~fb$^{-1}$, the goal for Run~IIb,
one hopes to achieve a precision of~\cite{snowmass}
\begin{equation}
\delta M_W=\pm 17~{\rm MeV}.
\end{equation}
For even larger data sets, the determination of the energy resolution
will be systematically limited by the uncertainty in the width of the
$Z$ boson~\cite{snowmass}.  This will restrict the accuracy to about 15~MeV
per channel per experiment.

At the LHC, for 10~fb$^{-1}$, one expects about $6\times 10^7$
$W\to\ell\nu$ ($\ell=e,\,\mu$) events, resulting in a statistical error
of $<2$~MeV~\cite{haywood}. If the lepton energy and momentum scales can
be determined with a precision of 0.02\% or better, the $W$ mass can be
measured with a total uncertainty of $\pm 15$~MeV~\cite{haywood}. 

\section{Detector-specific Uncertainties}
\label{sec:det_uncertainties}

Much of the understanding of experimental systematics comes from a 
detailed study of $Z$ bosons and hence as luminosity improves, 
systematic uncertainties should diminish in kind. Certainly, the 
scale and resolution of the recoil energy against the $W$ come from 
measurements of the $Z$ 
system. Likewise, background determination, underlying event studies, 
and selection biases depend critically, but not exclusively, on $Z$ 
boson data. Most importantly, the lepton energy and momentum scales 
depend solely on the $Z$ boson datasets. 
In the Run~I $W$ mass measurement, the largest systematic uncertainties
originated from the lepton energy and momentum scales, and the modeling
of the $W$ recoil. Non-$Z$ related recoil systematics were 
estimated to enter at the 10~MeV level. 

\begin{figure}[t!]
\centering\leavevmode
\includegraphics[width=3.25in]{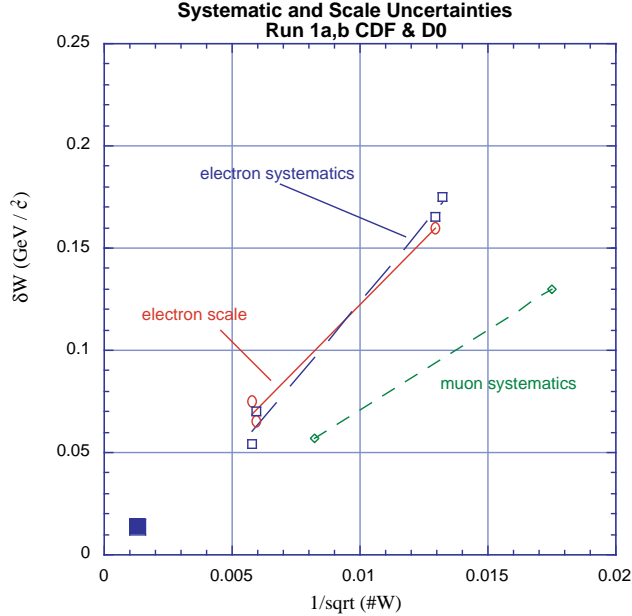}
\caption{Systematic uncertainties for each Run~I $M_W$ measurement. The
open squares are the four electron measurements from CDF and D\O, the
circles are the scale uncertainties from two D\O\ electron measurements
and the Run~Ib CDF measurement, and the diamonds are the systematic
uncertainties (excluding scale) for the CDF muon measurements. The large
box is the position of the extrapolated statistical uncertainties to the
Run~II luminosity. 
The lines are linear fits to each set of points.} \label{sys_emu}
\end{figure}

Figure~\ref{sys_emu} shows the CDF and D\O\ systematic uncertainties for
both electrons and muons as a function of $1/ \sqrt{N_W}$.
The calorimeter scale uncertainties for electrons is  currently tied to the
determination of fiducial di-lepton decay resonances, notably the $Z$
boson, but also the $J/\psi$, $\Upsilon$ and the $E/p$ dependence on the 
energy $E$, using electrons from $W$ and $Z$ decays. As the statistical
precision improves, the dominant feature of the scale determination
becomes its value in the region of $M_W$, so offsets and any low energy
nonlinearities become relatively less important and hence reliance on
the low mass resonances is reduced. On the other hand, for the muon momentum 
scale determination, where the observable is the curvature, low mass
resonances are also important. Figure~\ref{sys_emu} suggests that
this uncertainty is truly statistical in nature and extrapolates to
approximately the 15~MeV level. The ability to bound non-linearities 
using collider data may become a limiting source of error in Run~II. 

Figure~\ref{sys_emu} also shows the non-scale systematic 
uncertainties from both the CDF and D\O\ electron measurements of $M_W$
and the CDF muon measurement. Here the extrapolation is not as 
straightforward, but  there is clearly a distinct statistical 
nature to these errors. That they appear to extrapolate to negative 
values suggests that the systematic uncertainties may contain a statistically 
independent component for both the muon and the electron analyses. 

$Z\to\ell^+\ell^-$ data constrain
both the lepton scales and resolutions and an empirical model of the
hadronic recoil measurement. The determination of the lepton scales and
resolutions requires a measurement of the $Z$ boson mass, $M_Z$ and
width, $\Gamma_Z$, and 
calibration of these observables using the available LEP data. In order
to do this, it is necessary to use exactly the same theoretical input
which has been used to  
extract $M_Z$ and $\Gamma_Z$ at LEP, ie. to include the reducible and
irreducible ${\cal O}(g^4m^2_t/M_W^2)$ corrections to the effective
leptonic weak mixing parameter, $\sin^2\theta_{\rm eff}^l$, and the $W$ 
mass~\cite{dg}, in the calculation. A calculation of the complete ${\cal
O}(\alpha)$ corrections to $p\,p\hskip-7pt\hbox{$^{^{(\!-\!)
}}$} \rightarrow \gamma,\, Z \rightarrow \ell^+ \ell^-$ which also takes
into account the ${\cal O}(g^4m^2_t/M_W^2)$ corrections has recently
been completed~\cite{BBHSW}.

\section{Theoretical Uncertainties}
\label{sec:theo_uncertainties}

The $M_T$ lineshape simulation requires a theoretical model, as a 
function of the $W$ mass and width, of the differential cross
section. The $W$ mass is extracted by comparing data with templates of
the $M_T$ distribution for different values of $M_W$. The theoretical
model must take into account both QCD corrections and electroweak
corrections. 

QCD provides an understanding of the $p_T^W$ distribution and
thus of the hadronic recoil in $W$ events. For a quantitative
understanding of the $W$ (and $Z$) boson $p_T$ distribution it is
necessary to resum soft gluon radiation. In the low $p_T$ region, the
perturbative 
calculation must be augmented by a non-perturbative contribution which
depends on three parameters which are 
tuned to fit the $Z\to\ell^+\ell^-$ data.  
Electroweak radiative corrections, in particular final-state QED
radiation is important, because bremsstrahlung shifts the $W$ mass to
smaller values and affects the isolation
variables needed to select a clean $W$ sample.

To date, {\it ad hoc} event generators have been used in the $W$ mass
measurement. In Run~II, this measurement will reach a precision of tens
of MeV, requiring much more attention to detail in
Monte Carlo calculations.  Precision electroweak measurements in Run~II
should strive to use Monte Carlo programs that are common to both collider
experiments and include state-of-the-art QCD and electroweak
calculations. Currently, simulations are largely based on two programs:
{\tt RESBOS}~\cite{resbos} which includes resummed QCD corrections but
ignores electroweak radiative corrections, and {\tt WGRAD}~\cite{wgrad}
which includes ${\cal O}(\alpha)$ electroweak corrections, but ignores
QCD corrections. A merging of the two programs would be highly desirable
in order to achieve the goal of a unified program which includes both
QCD and electroweak radiative corrections.

The current $W$ production model has a number of shortcomings which lead
to systematic uncertainties in the $W$ mass measurement. These
uncertainties are discussed in the following.

\subsection{Parton Distribution Functions}

The transverse mass distribution is invariant under
the longitudinal boost of the $W$ boson.
However, the incomplete rapidity coverage of the detectors
introduces a dependence of the measured $M_T$ distribution on the
longitudinal momentum distribution of the produced $W$'s, determined by
the parton distribution functions (PDF's). Therefore, quantifying the
uncertainties in the PDFs and the resulting uncertainties in the $W$ mass
measurement is crucial. 

The measurement of the $W$ charge asymmetry at the Tevatron, 
which is sensitive to the ratio of $d$ to $u$ quark densities in the
proton, is of direct benefit in 
constraining PDF effects in the $W$ mass measurement. 
This has been demonstrated by the CDF experiment. From the variation
among the six reference PDFs,  
an uncertainty of 15~MeV was taken which is common to the 
electron and muon analyses. Since the Run~Ib charge asymmetry data is
dominated by statistical 
uncertainties, a smaller uncertainty is expected for the Run~II measurement.
Measurements of Drell-Yan production at the Tevatron can be used 
to get further constraints on PDFs.

Since the PDF uncertainty originates from the finite rapidity coverage of
the detectors, it is expected to decrease with the more complete 
rapidity coverage of the Run~II detectors.  
The advantage of a larger rapidity coverage has been demonstrated 
by the D\O\ experiment~\cite{largerap}: 
the uncertainty on the $W$ mass measurement using
their central calorimeter was 11~MeV, while
that using both the central and end calorimeter was 7~MeV.
With the upgraded calorimeters and trackers for the rapidity range $|\eta| > 
1$, the CDF experiment can measure the $W$ mass over a larger rapidity
range in Run~II.

There has been a systematic effort to map out the uncertainties
allowed by available experimental constraints, both on
the PDFs themselves and on physical observables derived from
them~\cite{giele,cteq}. 
This approach will provide a more reliable
estimate and may be the best course of action for precision
measurements such as the $W$ mass or the $W$ production cross section.

\subsection{$W$ Boson Transverse Momentum}
The neutrino transverse momentum is estimated by combining the measured 
lepton transverse momentum and the $W$ 
recoil:~${\vec p}_{T}(\nu)=-({\vec p}_{T}(\ell)+{\vec u})$.  It is clear
therefore that an understanding of both the underlying $W$ boson transverse 
momentum distribution and the corresponding detector response, usually 
called the recoil model, is crucial for a precision $W$ mass measurement. 
For the CDF Run~Ib $W$ mass measurement, the systematic uncertainties from 
these two sources were estimated to be $15-20$~${\rm MeV}$
and $35-40$~${\rm MeV}$, respectively, in each channel~\cite{cdf_prd}.

\begin{figure}[t!]
\centering\leavevmode
\includegraphics[width=3.25in]{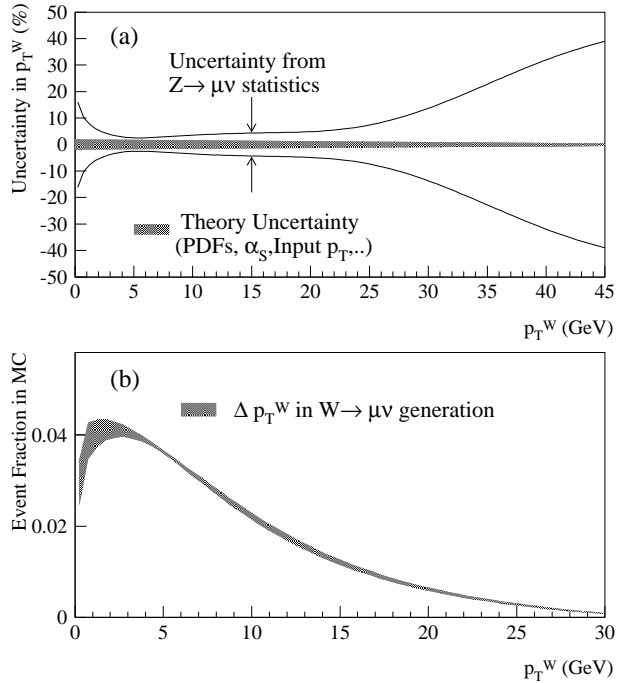}
\vspace{6.mm}
\caption{(a) A comparison of the two sources of uncertainty on the 
derived $p_{T}^{W}$ distribution. 
(b) The $p_{T}^{W}$ distribution extracted for the CDF Run~Ib 
$W\to\mu\nu$ mass measurement.}
\label{fig:wpt}
\end{figure}

The strategy employed in Run~I, which is expected to be used also in 
Run~II, is to extract
the underlying $p_{T}^{W}$ distribution from the measured $p_{T}^{Z}$ 
distribution ($Y$ is the weak boson rapidity):
\begin{equation}
\frac{d^{2}\sigma}{dp_{T}^{W}dY} = 
\frac{d^{2}\sigma}{dp_{T}^{Z}dY} \times 
\frac{d^{2}\sigma/dp_{T}^{W}dY}{d^{2}\sigma/dp_{T}^{Z}dY}\:, 
\label{eq:wpt_method}
\end{equation}
where the ratio of the $W$ and $Z$ differential distributions is 
obtained from theory. 
This method relies on the fact that the observed $p_{T}^{Z}$ 
distribution suffers relatively 
little from detector smearing effects, allowing fits to be performed 
for the true distribution. The experimental uncertainties, as in many
aspects of the $W$ mass measurement, are dominated 
by the available $Z\to \ell^{+}\ell^{-}$ statistics and should scale 
correspondingly with the delivered 
luminosity in Run~II. Theoretical uncertainties in the ratio of $W$ 
to $Z$ transverse momentum 
distributions contribute a further ${\cal O}(5)$~${\rm MeV}$ to the 
overall error. The two sources 
of uncertainty are compared for the CDF Run~Ib $W\to\mu\nu$ analysis 
in Fig.~\ref{fig:wpt}.

\subsection{QCD Higher Order Effects}

The charged leptons in $W\to\ell\nu$ are 
produced with an angular distribution determined by the
${\cal{O}}({{\alpha}_s^2})$ calculation of Ref.~\cite{MIRKES_HO_ANGLE} 
which, for $W^+$ bosons with a helicity of $-1$ with respect to the proton 
direction, has the form:
\begin{eqnarray}
 \frac{d\sigma}{d\cos{\theta_{\rm CS}}} \propto 
    1 + a_1(p_T^W)\cos{\theta_{\rm CS}} + a_2(p_T^W)\cos^2{\theta_{\rm 
CS}}.
   \label{EQUATION:ANG_DIST_W}
\end{eqnarray}
Here, $\theta_{\rm 
CS}$ is the polar direction of the charged lepton with respect to the 
proton direction in the Collins-Soper frame~\cite{COLLINS_SOPER_FRAME}. 
$a_1$ and $a_2$ are $p_T$ dependent parameters.
For $p_T^W$ = 0, $a_1 = 2$ and $a_2 = 1$, providing 
the angular distribution of a $W$ boson fully 
polarized along the proton direction.
For the $p_T^W$ values relevant to the $W$ mass analysis $(p_T^W < 
30$~GeV), the change
in $W$ polarization as $p_T^W$ increases only causes a modest change 
in the angular distribution of the decay leptons~\cite{MIRKES_HO_ANGLE}.

While the uncertainty associated with the change in the angular
distribution of the $W$ decay lepton due to higher order QCD corrections
(a few MeV) has been negligible for the Run~I measurements, 
it can not be ignored for the Run~II measurements.

\subsection{Electroweak Radiative Corrections}

The understanding of electroweak radiative corrections is crucial for 
precision $W$ mass measurements at the Tevatron. The dominant
process is final state photon radiation (FSR) from the charged lepton, the 
effect of which strongly depends on the lepton identification
criteria and the energy or momentum measurement methods employed. 
Calorimetric energy measurements, such as those employed in the 
electron channel, are more inclusive than track based momentum 
measurements used in the muon channel and the effect
of FSR is consequently reduced.  In the CDF Run~Ib $W$ mass 
measurement the mass shifts due to radiative effects were estimated 
to be $-65\pm 20$~${\rm MeV}$ and $-168\pm 10$~${\rm MeV}$ for the 
electron and muon channels, respectively~\cite{cdf_prd}. These effects
will be larger in Run~II due to increase in tracker material in CDF and
magnetic tracking in D\O. 

The Monte Carlo program used for the Run~I $W$ mass measurement 
incorporated a calculation of QED corrections by
Berends and Kleiss~\cite{berends}. This treatment, however, does not 
include initial state radiation (ISR), or soft and virtual
corrections. Furthermore, multi-photon radiation effects are ignored in
Ref.~\cite{berends}. The systematic uncertainties of the missing
radiative corrections were estimated to be $20$~${\rm MeV}$ and
$10$~${\rm MeV}$ in the electron and 
muon channels, respectively~\cite{cdf_prd}. Clearly these systematic 
uncertainties become much more significant in the 
context of statistical uncertainties of ${\cal O}(10)$~${\rm MeV}$ 
expected for $2$~${\rm fb^{-1}}$ in Run~II. 

In the last few years two independent calculations of the full ${\cal
O}(\alpha)$ electroweak (EW) corrections to $W$ production in Monte Carlo
form have been carried out~\cite{wgrad,DK,BW}. Calculating the EW
radiative corrections to resonant $W$ boson 
production, the problem arises how an unstable charged gauge boson can
be treated consistently in the framework of perturbation theory.
This problem has been studied in Ref.~\cite{dw} with 
particular emphasis on finding a gauge invariant decomposition of the 
EW ${\cal O}(\alpha)$ corrections into a QED-like and a modified weak
part. In Ref.~\cite{dw}, it was
demonstrated how gauge invariant contributions that contain the
infrared (IR) singular terms can be extracted from the virtual photonic 
corrections. These contributions can be combined with the also IR-singular 
real photon corrections in the soft photon region to form IR-finite 
gauge invariant QED-like contributions
corresponding to initial state, final state and interference
corrections. The collinear singularities associated with initial state
photon radiation can be removed by universal collinear counter terms
generated by ``renormalizing'' the parton distribution
functions, in complete analogy to gluon emission in QCD. 

\begin{figure}[t!]
\centering\leavevmode
\includegraphics[width=5.25in]{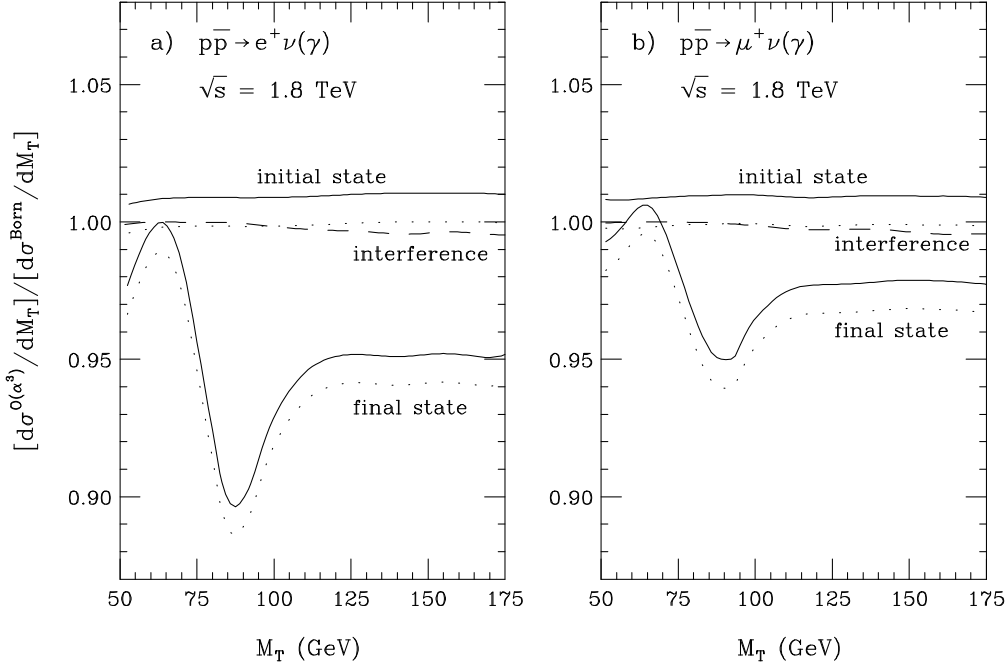}
\caption{Ratio of the ${\cal O}(\alpha^3)$ and lowest order 
cross sections as a function of the transverse mass for a) $p\bar p\to
e^+\nu(\gamma)$ and b) $p\bar p\to \mu^+\nu(\gamma)$ at
$\protect{\sqrt{s}=1.8}$~TeV for various individual contributions. The
upper (lower) solid lines show the result for the QED-like initial  
(final) state corrections. The upper (lower) dotted lines give the cross
section ratios if
both the QED-like and modified weak initial (final) state corrections are 
included. The dashed lines display the result if only the 
initial -- final state interference contributions are included. Lepton
identification requirements are not taken into account here.}
\label{fig:eight}
\end{figure}

The various individual
contributions to the electroweak
${\cal O}(\alpha)$ corrections on the $M_T$ distribution are shown
in Fig.~\ref{fig:eight}.
The initial state QED-like 
contribution uniformly increases  the cross section by about 1\% for 
electron (Fig.~\ref{fig:eight}a) and muon (Fig.~\ref{fig:eight}b) final 
states. It is largely canceled by the modified weak initial state 
contribution. The interference contribution is very small. The final
state QED-like contribution significantly changes 
the shape of the transverse mass distribution and reaches its maximum
effect in the region of the Jacobian peak, $M_T\approx M_W$.

Since final state photon radiation has a significant impact on the $W$
mass extracted from data, one has to worry about the effects of 
multi-photon radiation. A first attempt towards the inclusion of the
${\cal O}(\alpha^2)$ real photon corrections in $q\bar q'\to
W\to\ell\nu\gamma\gamma$ and $q\bar q\to Z\to\ell^+\ell^-\gamma\gamma$
was made in Ref.~\cite{stelzer}. More recently, two calculations of
higher order QED corrections to $W\to\ell\nu$ decays have been performed
which include both real 
and virtual corrections. Ref.~\cite{jadach} uses Yennie-Frautschi-Suura
exclusive exponentiation to calculate multi-photon radiation effects
whereas a QED parton shower approach is employed in
Ref.~\cite{nicrosini}. Using a simplified detector model, multi-photon
radiation in $W$ decays was found~\cite{nicrosini} to shift the $W$ mass
by a few MeV 
in the electron channel, and by about $10-15$~MeV in the muon
channel. In view of the expected precision of $15 - 30$~MeV for
$M_W$ in Run~II and at the LHC, it will be important to take
multi-photon radiation effects into account when extracting the $W$ mass
from data.

So far, no calculation exists which includes multi-photon radiation effects
in hadronic $Z$ boson production. In $Z\to\ell^+\ell^-$, both final
state leptons couple to photons, in contrast to $W\to\ell\nu$ where only
one of the final state particles carries electric charge. As a result,
one expects that multi-photon radiation affects the $Z$ boson mass and width
more strongly than the corresponding $W$ observables. Since $M_Z$ and
$\Gamma_Z$ are crucial in determining the lepton energy and momentum
scales, a calculation of $Z$ boson production in hadronic collisions
which takes into account higher order QED corrections is needed. 

\section{Other Methods of Determining $M_{W}$ at Hadron Colliders}
\label{sec:alternatives}

While the traditional transverse mass determination has 
been the optimal technique for the extraction of $M_{W}$
in previous hadron collider experiments,  other techniques may be 
employed in the future. These methods may shuffle or cancel
some of the systematic and statistical
uncertainties resulting in more precise measurements. 

\subsection{Transverse Momentum Fitting}
As noted in Sec.~\ref{sec:techniques}, the most obvious extensions of the 
traditional transverse mass approach to determining $M_W$ are fits of 
the Jacobian kinematical edge from the transverse momentum of both leptons. 
D\O\ has measured $M_W$ using all three distributions
and the uncertainties are indeed ordered as one would expect: The 
fractional uncertainties
on $M_W$ from the D\O\ Run~I measurements for the three methods of 
fitting are: 0.12\%
($M_T$), 0.15\% ($p_T(e)$), and 0.21\% ($p_T(\nu)$). As expected, the 
$p_{T}(e)$ method is slightly less precise than the measurement using
the transverse mass.  However, for a
central electron ($|\eta|\lessapprox 1$), the uncertainty in the $p_T(e)$ 
measurement due to the
$p_T^W$ model is 5~times that in the $M_T$ measurement. This is nearly
balanced by effects from  electron and  
hadron response and resolutions which are relatively worse for $M_T$. 
Accordingly, when there are sufficient statistics to enable cuts on 
the measured hadronic recoil, the measurement uncertainty from the 
$p_{T}^{W}$ model might be better controlled and enable the 
$p_{T}(e)$ 
measurement to compete favorably with the $M_{T}$ measurement which 
relies so heavily on modeling of the hadronic recoil. In order to 
optimize 
the advantages of all three measurements, the D\O\ final Run~I 
determination of $M_{W}$ combined the separate 
results~\cite{d0_measurements}.

The resolution sensitivity for muon measurements is even less than 
that for electrons so that has the benefit of slightly favoring a 
transverse mass measurement with muons over that for electrons.

\subsection{Ratio Method}
The $W$ mass can also be determined from the ratio of the transverse
mass distributions of the $W$ and $Z$ boson~\cite{MR,gike,shpakov}. 
The advantage of the method is that one can cancel common
scale factors in ratios  
and directly determine the quantity $r^{meas}\equiv {M_{W}\over M_{Z}}$, 
which can be compared with the precise value of $M_Z$ from the LEP
experiments. The downside of the ratio method is that the statistical
precision of the $Z$ sample is directly propagated 
into the resultant overall $\delta M_{W}$. 

D\O\ has carried out a ``proof of principle'' analysis, determining 
$M_{W}$ from the 
ratio of the transverse mass distributions of the $W$ and $Z$
boson~\cite{MR,gike,shpakov}. Since the $W$ and $Z$ boson mass are
different, the $Z$ boson data are rescaled to fit the $W$ boson
data. The preliminary D\O\ result for $M_W$ from Run~Ib data using the
transverse mass ratio method is~\cite{shpakov}
\begin{equation}
M_W = 80.115 \pm 0.211~{\rm(stat.)}\pm 0.050~{\rm (syst.)~GeV.}
\end{equation}
For comparison, the result from the fit to the $M_T$ distribution
is~\cite{d0_measurements} 
\begin{equation}
M_W = 80.440 \pm 0.070~{\rm(stat.)}\pm 0.096~{\rm (syst.)~GeV.}
\end{equation}
As expected, the transverse mass ratio method results in a much improved
systematic uncertainty. The statistical uncertainty, however, is
significantly larger than for the fit to the $M_T$ distribution. 

As mentioned in Sec.~\ref{sec:techniques}, systematic uncertainties
begin to limit the precision for $M_W$ which can be achieved from the
fit to the transverse mass distribution at the
Tevatron for integrated luminosities of 15~fb$^{-1}$ or more. Such high
luminosities may be achievable in Run~IIb. Up to
30~fb$^{-1}$ are conceivable if additional upgrades are made to the
accelerator complex~\cite{snowmass}. Table~\ref{tab:one} shows the
expected precision of the $W$ mass measurement from the transverse mass
fit and the $W$ to $Z$ transverse mass ratio, extrapolated from the
Run~Ib measurement by D\O~\cite{d0_measurements}.
\begin{table}
\caption{\label{tab:one} Projected uncertainties of the $W$ boson mass
measurement using the transverse mass fit and the transverse mass ratio
method for $W\to e\nu$.}
\begin{center}
\begin{tabular}{|cccc|cccc|}
\hline
 & & & & & & & \\[-3.mm]
$\int{\cal L}dt$ (fb$^{-1}$) & 2 & 15 & 30 & $\int{\cal L}dt$
(fb$^{-1}$) & 2 & 15 & 30\\[1.mm] 
\hline
 & & & & & & & \\[-3.mm]
\multicolumn{4}{|c|}{$M_T$ fit} & \multicolumn{4}{c|}{ratio method}\\[1.mm]
stat. uncert. (MeV) & 19 & 7 & 5 & stat. uncert. (MeV) & 44 & 16 & 11\\
syst. uncert. (MeV) & 19 & 16 & 15 & syst. uncert. (MeV) & 10 & 4 &
3\\[2.mm] 
\hline
 & & & &  & & & \\[-3.mm]
total uncert. (MeV) & 27 & 17 & 16 & total uncert. (MeV) & 45 & 16 &
11\\[2.mm]
\hline
\end{tabular}
\end{center}
\end{table}
For integrated luminosities larger than about 15~fb$^{-1}$ the
transverse mass ratio method gives an uncertainty for $M_W$ which is
smaller than that obtained with the traditional method of fitting the
$M_T$ distribution.

\section{Conclusions}
\label{sec:conclusions}

The measurements of the $W$ mass in Run~I of the Tevatron contribute
significantly to the world average of $M_W$. Close inspection of
the various systematic error sources  
leads us to believe that a $W$ mass measurement in Run~II at the 
$30$~${\rm MeV}$ level per experiment is achievable. Alternative methods
for determining $M_{W}$ at the Tevatron have been  discussed 
and may turn out to be more appropriate in the Run~IIb operating environment
than the traditional transverse mass fitting approach. Determination of
the $W$ mass 
at the LHC will be extremely challenging, using detectors that are not 
optimized for this measurement. Clearly, the $W$ mass 
measurement at the Tevatron in Run~II will contribute significantly
to a reduction of the overall uncertainty of $M_W$, and thus to a precise
indirect determination of the Higgs boson mass.

\section*{Acknowledgements}
The material presented in this report is largely based on the work of
the {\sl ``Precision Physics''} Working Group of the {\sl ``Run~II
Workshop on QCD and Weak Boson Physics''}, held at Fermilab in 1999, and
the work of the 2001 Snowmass {\sl ``Working Group on Precision
Electroweak Measurements''}. I
would like to thank the members of the working groups, in particular
R.~Brock, Y.K.~Kim and D.~Wackeroth for many useful discussions. I would
also like to thank the organizers of this workshop for the invitation
and financial 
support. This research was supported in part by the National Science
Foundation under grant No.~PHY-0139953.

\end{document}